\documentclass[sigplan,nonacm,screen]{acmart}
\setkeys{acmart.cls}{balance=false}
\setcopyright{none}
\acmConference[EuroSys '27]{European Conference on Computer Systems}{2027}{}
\usepackage{booktabs}
\usepackage{array}
\usepackage{xspace}
\usepackage{xurl}
\usepackage{listings}
\usepackage{tikz}
\usetikzlibrary{positioning,arrows.meta,calc}
\UseMicrotypeSet[protrusion]{basictext}

\makeatletter
\if@ACM@anonymous
  \newcommand{\proj}{x64spec\xspace}
\else
  \newcommand{\proj}{sail-x86\xspace}
\fi
\makeatother
\newcommand{\code}[1]{\texttt{#1}}
\newcolumntype{P}[1]{>{\raggedright\arraybackslash}p{#1}}

\begin{document}
\title{\proj: Formalizing x86-64 for Full-System Execution and Machine-Code Verification}
\author{Rui Ueyama}
\orcid{0009-0004-3216-9537}
\affiliation{ \institution{The University of Tokyo} \city{Tokyo} \country{Japan}}
\email{ruiu@is.s.u-tokyo.ac.jp}
\author{Takahiro Shinagawa}
\orcid{0000-0002-7016-7696}
\affiliation{ \institution{The University of Tokyo} \city{Tokyo} \country{Japan} }
\email{shina@is.s.u-tokyo.ac.jp}

\begin{abstract}
Executable formal specifications support reliable emulation, binary analysis, and formal verification of machine code.
However, existing x86-64 specifications remain incomplete, and direct translations of instruction pseudocode introduce data-dependent branches that increase path exploration during symbolic execution.
We present \proj, to our knowledge the most comprehensive executable formal specification of x86-64, written in Sail.
It covers 96\% (4,043 of 4,218) of the instructions in Intel's architecture manual and key system features including paging, exception handling, and virtualization.
We construct and validate \proj{} through large language model (LLM)-assisted translation, manual review, hardware differential testing, and full-system execution.
Its emulator boots unmodified Linux and Windows 95 and runs a guest using Linux KVM.
Validation also revealed a bug in Intel's architecture manual and a case where Linux relies on behavior that the manual leaves undefined.
To reduce path exploration during symbolic execution, we use equivalence-checked refactoring to reduce data-dependent branching in the specification.
To demonstrate the utility of \proj for machine-code verification, we extract 5,446 straight-line blocks from the handwritten assembly of seven widely used programs and ask an LLM to propose shorter rewrites. \proj proves 714 of the proposals equivalent to the originals and refutes 41 with concrete counterexamples.
\end{abstract}
\maketitle

\section{Introduction}
\label{sec:introduction}

Instruction set architectures (ISAs) define the behavior of machine instructions and architectural mechanisms visible to software.
Executable formal specifications express this behavior mathematically, enabling tools to simulate instruction execution and reason about machine code.
Such specifications provide a foundation for reliable emulation, binary analysis, and formal verification of machine code.
For example, proving an assembly-code rewrite correct requires accounting for its effects on registers, flags, and memory, including effects implicit in the instruction encoding.
For system software, such reasoning must also account for paging, exceptions, privilege transitions, and virtualization.

Arm and RISC-V already have executable formal specifications written in Sail~\cite{armstrong2019isa} that are sufficiently complete to boot Linux.
For x86-64, however, existing public formal specifications leave significant gaps in instruction coverage and system features.
The K semantics by Dasgupta et al.~\cite{dasgupta2019complete} cover only the user-mode instruction set.
The x86isa model, written in ACL2~\cite{goel2014simulation,goel2017engineering,goel2017books}, supports reasoning about both application and system programs, but it still lacks many instructions, most exception handling, and virtualization~\cite{x86isa2026source}.
An automatically translated Sail version of x86isa is also available~\cite{taylor2026sailx86}, though it inherits the architectural scope of the underlying ACL2 model.
Intel has also released a preview of an executable specification~\cite{intel2026specification}, but it omits system features, and modeling an OS boot remains an unofficial future goal that Intel expects will take ``a few more years'' to achieve~\cite{intel2026faq}.

Building a comprehensive specification that also supports formal analysis presents two challenges.
First, constructing and validating a faithful specification requires reconciling architectural descriptions and capturing interactions among architectural mechanisms.
An instruction's behavior depends on rules distributed across operation pseudocode, encoding tables, exception lists, and general descriptions elsewhere in the architecture manuals~\cite{intelsdm,amd64manual}.
These descriptions can be ambiguous or incomplete, and instruction-level tests can miss errors that arise only when paging, privilege checks, exception delivery, and execution modes interact.
Second, symbolic execution of the specification should scale to a broad range of instruction sequences, but a direct translation of instruction pseudocode introduces many data-dependent branches.
Because symbolic execution must explore feasible outcomes, these branches can multiply execution paths~\cite{kuznetsov2012efficient}, even when the machine code itself contains no conditional branches, making some instruction sequences expensive to analyze.

This paper presents \proj, to our knowledge the most comprehensive executable formal specification of x86-64 written in Sail.
It covers 96\% (4,043 of 4,218) of the instructions in Intel's Software Developer's Manual (SDM) and system features including real-mode execution, segmentation, paging, interrupts, exceptions, System Management Mode (SMM), and Virtual Machine Extensions (VMX), while leaving several deprecated, recent, or specialized extensions outside the current scope.
The same specification supports both full-system execution and symbolic equivalence checking of machine code.

To build confidence in \proj, we use a layered validation methodology that combines manual review, hardware differential testing, and full-system execution.
Because architectural rules are scattered across the Intel SDM, we manually review the large language model (LLM)-assisted translation against both instruction-specific definitions and shared architectural rules.
At the instruction level, we detect discrepancies in instruction behavior by running a large suite of automated differential tests on both Intel and AMD x86-64 processors, comparing register values, flags, memory effects, and exceptions.
At the system level, we exercise interactions among architectural mechanisms by running Linux, FreeBSD, MS-DOS, Windows 3.1, and Windows 95, together with a guest using Linux KVM.
Because the emulator generated by the official Sail compiler takes 681 seconds to boot Linux, we developed sail-llvm, an LLVM-based Sail compiler that reduces the boot time to 33 seconds and enables repeated full-system testing during development.
These full-system executions exposed various errors that had escaped instruction-level testing, including errors in exception delivery, paging, segmentation, and privilege transitions.

To address the symbolic-execution challenge, we refactor the specification by replacing many data-dependent branches in Sail instruction definitions with branchless bit operations.
Because the resulting code can differ substantially from the manual-oriented version, we use Sail's SMT backend to prove that each refactoring computes the same result as the code it replaces.
This lets us validate a specification that closely follows the manual before transforming it into a form that reduces path exploration during symbolic execution.
Across 6,192 straight-line blocks extracted from handwritten assembly in seven widely used programs, the refactored model completes symbolic execution within 60 seconds for 92\% of the blocks, compared with 76\% for the original model using Isla's linearization.

To demonstrate the utility of \proj{} for machine-code verification, we apply it to handwritten x86-64 assembly from seven widely used programs.
For the 5,446 blocks that the refactored model can symbolically execute on a single path, we ask an LLM to propose shorter rewrites and check every proposal for equivalence.
Of 789 proposed rewrites, we prove 714 equivalent to the originals and refute 41 with concrete counterexamples; 29 remain unresolved, and 5 do not assemble.
These results show that a comprehensive executable ISA specification can support formal reasoning about handwritten assembly from widely used software.

Our contributions are:
\begin{itemize}
\item
A comprehensive executable specification of x86-64 in Sail, covering a broad range of instructions and system features sufficient for full-system execution, including operating-system boot and virtualization.
\item
A layered validation methodology combining manual review, hardware differential testing, and full-system execution to uncover specification errors missed by instruction-level testing and discrepancies among architectural documentation, hardware behavior, and software dependencies.
\item
An equivalence-checked refactoring of the specification that substantially reduces path exploration during symbolic execution, together with a large-scale application to checking proposed rewrites of handwritten x86-64 assembly from widely used software.
\end{itemize}

\section{Sail and Its Tools}
\label{sec:sail}

Sail is a language for writing executable formal specifications
of instruction set architectures~\cite{armstrong2019isa}. Its
syntax resembles that of the pseudocode used in ISA manuals: instruction
behavior is expressed with assignments, conditionals, loops, and
function calls. A model typically declares the processor's registers as
global state and defines functions that read and update that
state.

For example, \code{POPCNT} is an instruction that counts the bits
set to one in its source operand. The SDM describes its operation
with the following pseudocode:

\par\noindent\begin{minipage}{\linewidth}
\begin{lstlisting}
    Count = 0;
    For (i=0; i < OperandSize; i++) {
        If (SRC[i] = 1) THEN Count++; FI;
    }
    DEST := Count;
\end{lstlisting}
\end{minipage}\par

For a 64-bit source operand, the same calculation can be written
in Sail as follows:

\par\noindent\begin{minipage}{\linewidth}
\begin{lstlisting}
    val popcnt64 : bits(64) -> int
    function popcnt64(src) = {
      var count : int = 0;
      foreach (i from 0 to 63) {
        if src[i] == 0b1 then
          count = count + 1
      };
      count
    }
\end{lstlisting}
\end{minipage}\par

The \code{val} line declares the function's type. Sail is a typed
language: registers are usually represented as bitvectors, and
integers are often used for intermediate calculations. Here,
\code{bits(64) -> int} means that the function takes a 64-bit
bitvector and returns an integer.

The body follows the manual's loop, accumulating the count in a
mutable local variable declared with \code{var}. The last
expression, \code{count}, is the function's return value. An
instruction definition for \code{POPCNT} can call this function,
write the result to its destination, and update the flags.

Pseudocode in ISA manuals often uses exceptions. For example,
the SDM's \code{DIV} pseudocode begins with
\code{IF SRC = 0 THEN \#DE; FI;}, raising a divide-error exception
if the divisor is zero. Sail has \code{throw} and
\code{try}/\code{catch}. In \proj, the code for \code{DIV} may
throw a value that represents \code{\#DE}. This transfers control
to the catch clause we wrote in the model's main loop, which in
turn invokes the exception-delivery code.

The same specification in Sail can be used both to execute machine code
and to reason about it. The Sail compiler can translate a model
into C++ code for an instruction-set simulator. Combined with
memory and device models, it can run existing software, including
operating systems. This
also makes the specification usable as a reference implementation
for testing.

Symbolic execution represents unknown inputs as variables and
computes expressions describing how the outputs depend on them.
For example, let $x$ and $y$ represent the 64-bit values
in \code{RAX} and \code{RBX}. Symbolically executing
\code{XOR RAX, RBX} gives the expression $x \oplus y$ for the new
value of \code{RAX}, where $\oplus$ denotes bitwise XOR.

Isla~\cite{armstrong2021isla} provides symbolic execution for
Sail models. Given a model and a machine-code sequence, it uses
the model's instruction definitions to compute expressions for
the values of the registers and memory after the sequence. When a
branch depends on a
symbolic input, Isla follows both outcomes as separate paths and
records, for each, the condition under which it is taken. The
\code{POPCNT} loop branches on one bit of the source in every
iteration, so with a symbolic source the number of paths doubles
64 times.

A satisfiability modulo theories (SMT) solver decides whether a
logical formula has a solution. We use it to check whether two
instruction sequences are equivalent. Isla executes both from the
same symbolic initial state, and from the resulting symbolic
expressions and path conditions, we build a formula that says the
two sequences disagree on some output. A solution of the formula
is a counterexample, an input on which they differ. If the solver
proves that no solution exists, the sequences are equivalent for
every input. The solver may also run out of time, which leaves the
question open. Isla itself also calls an SMT solver during
symbolic execution, to discard paths whose conditions cannot be
satisfied.

Sail's own SMT backend provides another route to the solver. It
translates properties of Sail functions directly into formulas,
without executing a machine-code sequence through Isla. For
example, it can check whether two Sail functions return the same
result for every input. We use this
backend to verify refactorings of our model itself.

\section{Overview of \proj}
\label{sec:construction}

This section describes the scope, development process, and structure of \proj.
Development combined LLM-assisted translation of the SDM into Sail, testing, and equivalence-checked refactoring.
Sections~\ref{sec:validation}--\ref{sec:manual} present the validation, the refactoring for symbolic execution, its application to assembly rewrite verification, and the architectural findings revealed by validation.

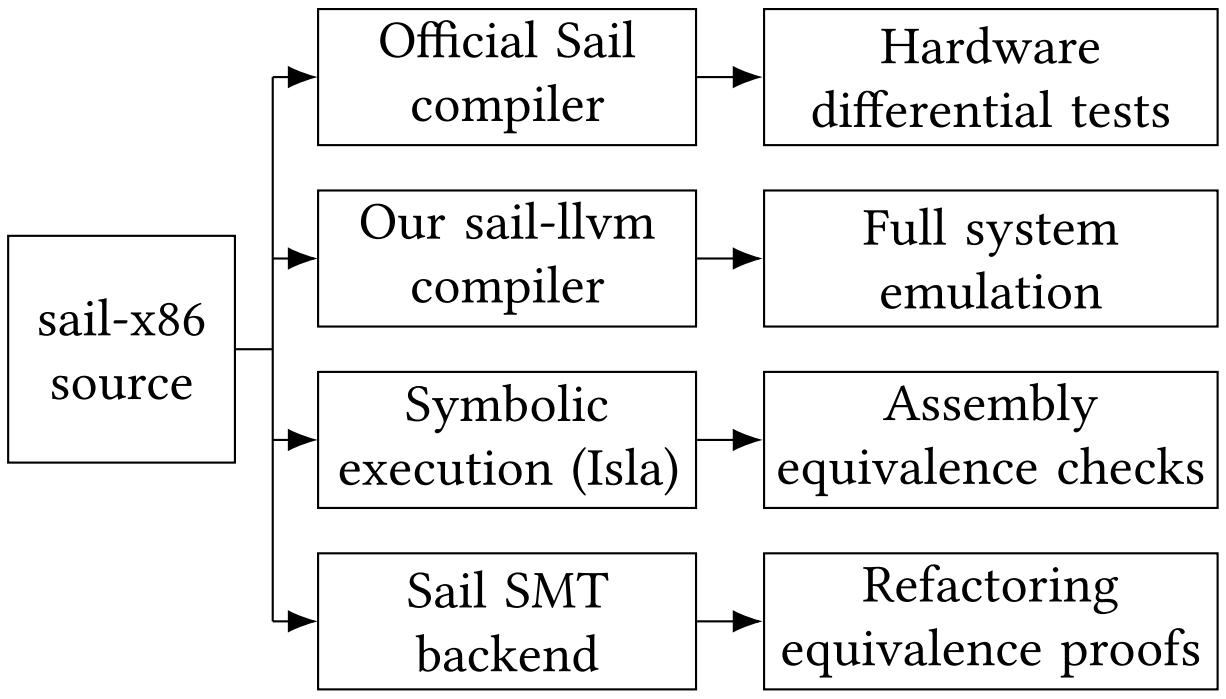
\begin{figure}[t]
\centering
\begin{tikzpicture}[
  box/.style={draw,align=center,minimum height=9mm,inner sep=0pt},
  arr/.style={-{Latex[length=2mm]}}]
\node[box,text width=25mm] (t1) at (2.55,0)    {Official Sail\\compiler};
\node[box,text width=25mm] (t2) at (2.55,-1.2) {Our sail-llvm\\compiler};
\node[box,text width=25mm] (t3) at (2.55,-2.4) {Symbolic\\execution (Isla)};
\node[box,text width=25mm] (t4) at (2.55,-3.6) {Sail SMT\\backend};
\node[box,text width=30mm] (u1) at (5.75,0)    {Hardware\\differential tests};
\node[box,text width=30mm] (u2) at (5.75,-1.2) {Full system\\emulation};
\node[box,text width=30mm] (u3) at (5.75,-2.4) {Assembly\\equivalence checks};
\node[box,text width=30mm] (u4) at (5.75,-3.6) {Refactoring\\equivalence proofs};
\node[box,text width=15mm,minimum height=15mm] (src) at (0,-1.8) {\proj\\source};
\draw (src.east) -- (1.0,-1.8);
\draw (1.0,0) -- (1.0,-3.6);
\foreach \y in {0,-1.2,-2.4,-3.6} { \draw[arr] (1.0,\y) -- (1.3,\y); }
\foreach \i in {1,2,3,4} { \draw[arr] (t\i.east) -- (u\i.west); }
\end{tikzpicture}
\caption{Four uses of the same Sail source. Both compilers produce
the same emulator; the official one is used for the hardware
comparison and ours where speed matters. Isla and the SMT backend
turn the definitions into queries for an SMT solver.}
\label{fig:uses}
\Description{The Sail source feeds two compilers that produce one
emulator, used for hardware differential tests and full system
emulation, and it feeds Isla and the Sail SMT backend, used for assembly
equivalence checks and refactoring equivalence proofs.}
\end{figure}

Figure~\ref{fig:uses} shows how we use the same model for execution
and formal reasoning. We use the official Sail compiler
to generate the emulator for hardware differential
testing, and our own compiler, sail-llvm, for faster system
emulation (\S\ref{sec:compiler}). Isla checks assembly
transformations (\S\ref{sec:applications}), while Sail's SMT
backend checks equivalence between original and rewritten Sail
functions (\S\ref{sec:symbolic}).

\subsection{Architectural Scope}
\label{sec:scope}

\proj aims to cover every instruction the SDM defines. No
hardware implements them all, however: some instructions are
deprecated, others exist only on the newest processors. We treat
the hardware as the oracle for what an instruction actually does,
to make sure that we have not misunderstood its behavior.
We therefore limited ourselves to the instructions we can test
using our own machines, and among those we gave priority to
what ordinary operating systems and applications use. This is why
SGX, SMX, MPX, CET, TSX, and Key Locker are not yet supported,
though we expect adding them to be more work of the same kind
rather than a new problem.

Table~\ref{tab:scope} compares the instruction groups and system
mechanisms supported by K, x86isa, and \proj. \proj covers more of the
architecture than K and x86isa, enough to run the whole stack, from
a BIOS through an operating system such as Linux to a virtual
machine inside it.

\begin{table*}[t]
\caption{Architectural scope of existing models and \proj. A check
mark means that the instruction group or mechanism is part of the
formal semantics, ``partial'' means that only part of it is, and a
dash means that none of it is. The AVX-512 column excludes
instructions specific to Xeon Phi.}
\label{tab:scope}
\centering
\setlength{\tabcolsep}{3pt}
\begin{tabular}{@{}lcccccccccc@{}}
\toprule
Model & x87/MMX & SSE & AVX/AVX2 & AVX-512 &
Real mode & Segments & Paging & Exceptions & SMM & VMX \\
\midrule
K~\cite{dasgupta2019complete} &
-- & partial & partial & -- &
-- & -- & -- & -- & -- & -- \\
\addlinespace[3pt]
x86isa~\cite{x86isa2026source} &
partial & partial & partial & -- &
-- & \checkmark & \checkmark & partial & -- & -- \\
\addlinespace[3pt]
\proj &
\checkmark & \checkmark & \checkmark & \checkmark &
\checkmark & \checkmark & \checkmark & \checkmark & \checkmark & \checkmark \\
\bottomrule
\end{tabular}
\end{table*}

Table~\ref{tab:coverage} compares the number of supported
instructions among the three models. We count each row of an
opcode table in SDM Volume~2 as one instruction, distinguishing
mnemonics, operand widths, and legacy, VEX, and EVEX encodings.
This gives a total of 4,218 instructions in the February 2026
edition, the one we worked from. We determined
support by inspecting each model's decoder and instruction
definitions; for K, which has no decoder, we checked the
corresponding assembly semantics.

\begin{table}[t]
\caption{SDM instructions supported by each model, out of 4,218.}
\label{tab:coverage}
\centering
\begin{tabular}{@{}lr@{}}
\toprule
Model & Instructions \\
\midrule
K & 1,632 (39\%) \\
x86isa & 1,183 (28\%) \\
\proj & 4,043 (96\%) \\
\bottomrule
\end{tabular}
\end{table}

Of the 175 unsupported instructions, 56 belong to the extensions listed
above. Fourteen are AVX-512 instructions specific to the discontinued
Xeon Phi processors. Most of the other 105 require newer
processors or features.

Besides the missing instructions, our model inherits two further limits
from Sail. A Sail specification
describes a single processor and leaves
memory ordering to a separate concurrency model, which we have not
written. Floating-point arithmetic is not built into the Sail
language, and existing models delegate it to external C/C++
functions. We follow suit. Therefore, multiprocessor execution and
floating-point arithmetic lie outside our formal semantics.

\subsection{Development Process}
\label{sec:process}

We used Claude Code, principally with Claude Opus 4.6, to
translate the SDM into Sail. We first asked the LLM to build a
checklist from the SDM for each group of instructions, such as
basic arithmetic or SSE, and for the general rules that apply
across instructions, such as operand sizes, addressing modes,
prefix legality, exception conditions, and flag behavior. The
checklists kept the remaining tasks visible to both the developer
and the LLM. Without them, the LLM often lost track of what it had
already implemented and what remained.

Work proceeded one group at a time. The LLM implemented a group
of instructions or a system mechanism from the manual, wrote
hardware differential tests for it, and fixed what they caught.
The developer reviewed the result and chose the next group.

A faithful translation of the SDM into Sail is verbose and
repetitive. This is because the SDM gives pseudocode separately
for each instruction, even when several instructions share the same
structure. For example, the SDM writes out the same element loop for
\code{PADDB}, \code{PADDW}, \code{PADDD}, and \code{PADDQ}, and
again for each of them with 64-, 128-, 256-, and 512-bit operands.
Subtraction, comparison, minimum, maximum, and the logical
operations repeat the pattern with their own operators. Our goal is to create a Sail model that is not only
machine-readable but also human-readable, so we refactored the code
to reduce repetition. The hardware differential tests written for
each instruction served as the regression suite for this refactoring.
As a result, the model shrank from about 45,000 lines to 23,000
before further additions brought it to about 28,000.
We also removed data dependent branches from our Sail code so that
symbolic execution finishes within a time limit for more code
(\S\ref{sec:symbolic}).

Sail describes only the processor. To run an operating system with
the emulator generated from the model, we also have to emulate
physical memory and peripherals such as an interrupt controller.
We had the LLM write this code too, in C++
(\S\ref{sec:validation-results}).

Most of the work on the model took place in about four weeks.
Within the first week, the model covered most of the instruction
set and 64-bit system mode, and Linux booted to an initramfs
shell. At that point the emulator loaded the kernel directly
into memory and started it in 64-bit mode, since real and
protected modes were not yet supported. By the third week, the
model had gained real mode and SMM, the emulator had gained
support for common PC peripherals such as a hard drive and VGA
text mode, and FreeDOS booted from SeaBIOS. VMX followed in the
fourth week, and Linux inside the emulator successfully ran a KVM
guest.

\subsection{Structure of the Model}
\label{sec:model}

The model is about 28,000 lines of Sail in 50 files. One file
declares the processor state: the instruction pointer (RIP), sixteen
64-bit general-purpose registers (GPRs), thirty-two 512-bit SIMD
registers, the flags, the segment registers with their descriptor
caches, control and model-specific registers, x87 state, and
virtualization state.
GPRs are read and written through accessors that implement
the architecture's width rules, so that a 32-bit write to a
GPR clears the upper half while an 8-bit or
16-bit write preserves the other bits.

The model's main function is \code{step}, which executes one
instruction. It first delivers any pending interrupt, then reads
the instruction at \code{RIP} and executes it, and, if the
instruction raised an exception, delivers it. The emulator calls
\code{step} in a loop.

Decoding x86-64 instructions is hard. An instruction is up to 15
bytes long, can carry several prefixes, and cannot be delimited
without reading it byte by byte.
Our model reads one byte at a time and dispatches
on its value, and once the bytes identify an instruction, the
code at that point performs the instruction's operation. This
differs from the Sail model of RISC-V, which decodes an instruction
into a data structure and executes it in a separate step.
We did not separate the two because x86-64 instructions are so
irregular that a representation of decoded instructions would have
to be as irregular as the instructions it represents, defeating its
purpose.

The SDM leaves some result bits undefined. Their values on real
hardware can differ between processors, and even between P-cores
and E-cores within the same processor~\cite{yedidia2025deterministic}.
We follow the SDM by representing these bits with Sail's
\code{undefined} construct, which denotes an unspecified value.
We discuss an exception in \S\ref{sec:bsf}.

\section{Validating \proj}
\label{sec:validation}

We validated \proj{} by comparing instruction results with hardware
and running operating systems on the emulator. We used both methods
throughout development. This section describes the two methods,
what each of them found, and the compiler we built to make the
operating-system runs fast enough to repeat.

\subsection{Hardware Differential Testing}

In a hardware differential test, we write instruction bytes to
memory, run them from the same initial machine state on the
emulator and on a host CPU through KVM, and compare the resulting
states. We use the hardware as the oracle rather than writing the
expected result into each test, because a test written by an LLM
that misunderstood an instruction would only check the model
against the same misunderstanding.

The suite follows three rules. First, when the LLM adds an instruction to
the model, it also writes and commits a test that runs the
instruction on the hardware and the emulator and checks that the
two agree. Second, since an instruction cannot be tested on all
possible inputs, source operands are set to predetermined boundary
values, chosen to expose a bug if there is one. Third, exception
conditions are tested as cases of their own, with the same weight
as normal execution.

For each arithmetic and logic instruction, the test harness
initializes the instruction's source operands with values such as
0, 1, $-1$, \code{0x5555}\ldots, \code{0x1234}\ldots, and the
signed and unsigned limits of 8, 16, 32, and 64-bit values and
their neighbors, thirty patterns in all, and then runs the
instruction. An instruction with two source operands, such as
\code{ADD}, therefore runs $30 \times 30 = 900$ times at each of
its four operand widths. Instructions that read the carry flag
run with it clear and set. Floating-point instructions run on
special values such as signed zeros, infinities, quiet and signaling NaNs,
and denormals, under each rounding mode. For AVX instructions,
what varies is the encoding rather than the data: every instruction
runs at the three vector lengths, with and without masking and
zeroing, and with register, memory, and broadcast operands. We test
each of the other instructions with one or a few fixed inputs,
enough to check that it decodes and computes as the manual says.
We also test instructions under conditions that cause processor
exceptions, such as division by zero. We compare the exception type
and, where applicable, the error code and faulting address between
the hardware and the emulator.

After a run of each instruction, the registers and flags of the
hardware are compared with those of the emulator, and any
difference is treated as a bug in the model, except in the bits
that the model marked as \code{undefined}, which are not compared.

We run each test twice, filling the GPRs, SIMD registers, mask
registers, and flags with all zeros or all ones before setting the
test's inputs. This helps detect both unintended changes to bits
that should be preserved and missing writes hidden by an initial
value that happens to match the expected result. In all, the suite
runs 139,622 tests on an AMD Ryzen Threadripper 7980X and 142,006
on an Intel Xeon Platinum 8562Y+. The Xeon runs more because it
supports extensions the Threadripper lacks, such as AVX-512 FP16
and AMX.

Floating-point arithmetic is computed by external C/C++
functions rather than by Sail code (\S\ref{sec:scope}), so a
matching result does not validate the arithmetic in the model. We
still compare these results, with a numerical tolerance for
approximate instructions, because a result far from the
processor's would reveal an error elsewhere, such as an operand
read from the wrong place.

KVM runs each test in 64-bit mode at supervisor privilege, with
the first 2 MB of the address space identity mapped to physical
memory. This environment suffices for most instructions, because
their behavior does not depend on the state around them. Tests of
32-bit code switch the guest to compatibility mode.

Although every instruction was added to the model together with a
test, a test can have an oversight that leaves some of the
instruction's behavior untested. We therefore measured the suite's
code coverage of the Sail model with the Sail compiler's
\code{--c-coverage} option and asked the LLM to raise it. The
tests it wrote found several errors. \code{CMPXCHG}, for example,
compares its destination with the accumulator \code{RAX} and, on a
mismatch, copies the destination into \code{RAX}. The model copied
on a match as well, which is harmless unless \code{RAX} is itself
the destination, where the copy undoes the exchange just made. No
earlier test had used \code{RAX} as the destination. Raising the coverage of the
decoder also found instructions we had missed, encodings for which
the emulator raised an undefined-opcode exception where the
hardware did not (\S\ref{sec:x87-aliases}).

Instructions that change the execution mode itself, and system
mechanisms such as paging, are tested by the method of the next
subsection.

\subsection{System-Level Testing}
\label{sec:validation-results}

A hardware differential test runs a single instruction in a fixed
environment and checks the result each time, so errors that depend
on a combination of states may pass through it. Mode transitions,
paging, segmentation, and virtualization cannot be tested with a
single instruction, because their effects show only in the
instructions that follow. To test them, we tried to boot various
operating systems on the emulator generated from \proj, fixing
errors in the model as each attempt exposed them.

For system-level testing, any operating system would work in
principle as long as it runs on an x86-64 processor. We did not,
however, want systems that require many kinds of peripherals. A
Sail model describes only the processor, and the more C++ we write
for PC peripherals, the more system-level testing becomes a test
of that C++ rather than of the model. We therefore limited the
platform to what SeaBIOS needs plus a few devices: physical memory,
the 8259 interrupt controllers, the 8254 timer, the CMOS clock, the
8042 keyboard controller, a 16550 serial port, IDE disks and an
ATAPI CD-ROM, a floppy controller, a DMA controller, VGA text mode
and a VBE framebuffer, a local and an I/O APIC, and a minimal PCI
host bridge. That came to about 5,000 lines of C++, which the LLM
wrote. The platform has no UEFI firmware, so Windows 11 is absent from
Table~\ref{tab:validation}, which lists the systems we ran.

Each system covers a different part of the architecture. Linux
takes the processor from 16-bit real mode through protected mode
to long mode and exercises interrupts and paging, including
5-level paging. Inside the same Linux, a test program uses the
unmodified \code{kvm-intel} driver to run a guest inside the emulated
machine, which is our only test of VMX: VM entry and exit, guest
paging, and Extended Page Tables. MS-DOS and FreeDOS boot through
SeaBIOS and run in real mode, and since SeaBIOS calls its 32-bit
code through System Management Mode, booting them also tests SMM.
Windows 3.1 and Windows 95 run their DOS prompts in virtual-8086
mode. FreeBSD, Haiku, xv6, and memtest86+ exercise the same
mechanisms with different code.

\begin{table*}[t]
\caption{Results of booting various operating systems and running
system software on the emulator generated from \proj.}
\label{tab:validation}
\centering
\begin{tabular}{@{}P{.24\textwidth}P{.72\textwidth}@{}}
\toprule
Workload & Result \\
\midrule
FreeBSD 14.5 & Loader and 64-bit kernel booted from the CD; single-user shell reached \\
FreeDOS 1.3 & DOS command prompt reached \\
Haiku R1 beta 5 & Boot loader and 64-bit kernel started; boot screen displayed \\
Linux 6.19.6 & Shell reached from 16-bit entry; a 32-bit program
also ran in compatibility mode \\
Linux 6.19.6, 32-bit kernel & Shell reached from 16-bit entry \\
Linux \code{kvm-intel} driver & Guest paging, memory, I/O, and halt
checks passed \\
memtest86+ 7.20 & Boots, detects the CPU and
memory, and runs its first tests with no error reported \\
MS-DOS 6.22 & Boots to the command prompt; basic commands work \\
Windows 3.1 on MS-DOS 6.22 & Boots to the
Program Manager desktop in 386 enhanced mode; a DOS session
runs in virtual-8086 mode \\
Windows 95 & Boots to the desktop with a
working Start menu; an MS-DOS window runs in virtual-8086 mode;
clean shutdown and reboot \\
xv6 (2020-08-10) & Shell reached \\
\bottomrule
\end{tabular}
\end{table*}

System-level testing found various errors in the model that
hardware differential testing had not revealed. These included:
\begin{itemize}
\item Privilege and segmentation. The instruction tests run at
  ring 0 with flat segments, so anything that depends on the
  privilege of an implicit access or on a descriptor's contents
  never came up, such as a descriptor cache left stale after a
  fast system call.
\item Exception and interrupt delivery. The tests provoke faults
  but never deliver an interrupt through a gate to a handler, so
  the error code, the gate's privilege, and the interrupt shadow
  after \code{STI} went unchecked.
\item Paging. The tests use one identity-mapped region, so no
  access ever crossed a page or faulted halfway through. The model
  translated only the first address of an access that crossed a
  page boundary and continued through consecutive physical bytes,
  which corrupted memory once Linux mapped adjacent virtual pages
  to distant physical ones.
\item Mode-dependent decoding. Nearly all tests run in 64-bit
  mode, so behavior that changes with the mode or the operand size
  went untested. The bytes C4 and C5, for example, begin a VEX
  prefix in 64-bit code but encode \code{LES} and \code{LDS} in
  32-bit code unless the next byte rules that out, and the model
  took them for VEX prefixes everywhere.
\end{itemize}

\subsection{Speeding Up System Emulation}
\label{sec:compiler}

The official Sail compiler is slow, and so is the emulator it
generates. Compiling our model to C++ takes about 90 seconds, and
the emulator needs about 11 minutes to boot Linux to a shell.
Every test that runs on Linux, such as the KVM test, therefore
begins by waiting 11 minutes for Linux to boot, so trying even a
one-line change to the model takes more than ten minutes. That
pace was too slow to work at, so we decided to write our own Sail
compiler, sail-llvm, from scratch, with speed as its goal.

The authors of Sail and of Pydrofoil identify arithmetic on GMP,
an arbitrary precision integer library~\cite{gmp}, as a major
source of the emulator's execution
time~\cite{armstrong2019isa,bolz2025pydrofoil}. Bitvectors and
integers are the main primitive types of Sail. The official
compiler maps them to \code{int64\_t} or \code{uint64\_t} when it
can determine that a value fits in 64 bits. Otherwise it maps them
to heap-allocated GMP values, on which even a simple operation
such as an addition becomes a GMP function call.
Most values in our model are compiled to GMP types.
We therefore focused on eliminating this cost.

In sail-llvm, we compile both bitvectors and
Sail integers into LLVM virtual registers. Values in virtual
registers are plain integers rather than opaque GMP objects, so
LLVM's optimization passes can apply ordinary compiler
optimizations to them, such as constant propagation and register allocation.
This design raised two challenges. First, the width of a Sail
bitvector is not always known at compile time, as a Sail function can
take a polymorphic bitvector of any width as an argument. Second,
Sail integers have unbounded precision.

We solved them as follows. For the first, sail-llvm treats
a function that takes a bitvector of variable width as a template
rather than as a concrete function, and instantiates a separate
function for each concrete width that appears at its call sites.
In our model, this removes all polymorphic bitvectors. For the
second, we simply assume that integers fit in 128 bits and compile
them to LLVM's \code{i128} type. In
Sail, large values such as SIMD registers are represented
as bitvectors, and integers serve as temporaries for computation.
In our model, those temporaries fit in 128 bits.
With these changes, sail-llvm compiled our model in 23
seconds instead of the official compiler's 90, and its emulator
booted Linux to a shell in 33 seconds instead of 681.

Most system-level testing ran on the emulator that sail-llvm
generated. Once an operating system in
Section~\ref{sec:validation-results} booted on that emulator, we
recompiled the same model with the official Sail compiler and
checked that the system behaved the same. This gave us the speed of
sail-llvm while testing the model and confirmed that the two
compilers agree.

\section{Making Symbolic Execution Practical}
\label{sec:symbolic}

When we began using \proj for symbolic execution, we noticed that even
short instruction sequences took a long time to execute. The cause was
the large number of data dependent branches in the Sail code. The
SDM's pseudocode often uses conditionals to describe instruction
behavior. For example, the pseudocode for \code{CMOVA} (conditional
move if above) uses an \code{if} to decide whether to copy a value
based on the condition flags. A straightforward translation preserves
these branches, so Isla may have to explore both sides of each one
for symbolic execution. The number of execution paths can grow
exponentially with the number of such branches. The problem is
especially severe for SIMD instructions.
Even a single SIMD instruction with sixteen lanes can produce
$2^{16}$ paths if its pseudocode has one data dependent branch
per lane.

Replacing branches with dataflow operations is an established
technique for reducing path exploration in symbolic
execution~\cite{collingbourne2011symbolic,wagner2013overify}.
Isla can do this itself, in a pass it calls linearization, but
only for functions that are pure and free of loops, and only for
the functions it is told to linearize. We therefore rewrote the
model's Sail source to remove as many data dependent branches as
possible from the instruction definitions.

\subsection{Removing Data Dependent Branches}

Consider the SSE2 instruction \code{PSUBUSB}, which performs unsigned
saturating subtraction for each 8-bit lane. Our original Sail model
for the instruction contained a branch for each lane,
\code{if x < y then 0 else x - y}, because it was a straightforward
translation of the SDM's pseudocode.

We can, however, compute the same subtraction in a branchless manner
using a well known masking technique as described in Hacker's
Delight~\cite[\S2--19]{warren2013hackers}. Here is how: let $b$ be
the borrow bit of $x-y$, and let $m$ be the byte formed by repeating
$b$ eight times. Then
$\max(x-y,0)=(x-y)\mathbin{\&}({\sim}m)$, where $\&$ and $\sim$ denote
bitwise AND and NOT. This works because $b=1$ when $x<y$, making
${\sim}m$ all zeros.
Otherwise, ${\sim}m$ is all ones and passes the subtraction result through.
In Sail, this is:

\par\noindent\begin{minipage}{\linewidth}
\begin{lstlisting}
    let d : bits(9) = (0b0 @ x) - (0b0 @ y);
    let m : bits(8) = replicate_bits(d[8], 8);
    d[7..0] & ~(m)
\end{lstlisting}
\end{minipage}\par
Here \code{@} concatenates bitvectors. The first line extends
\code{x} and \code{y} to nine bits by prepending a zero bit to each.
The most significant bit of \code{d} is the borrow bit. Repeating
it eight times gives the desired mask.

We first translated the SDM faithfully into Sail and then asked
the LLM to remove as many branches as it could.
The branches that remain in the instruction definitions after that are, with
few exceptions, ones the manual itself makes on the machine state:
the checks that raise exceptions, control transfers, masked loads
and stores, and reads of system state such as the XSAVE
components. Ten branches on operand values remain, such as the
conditional write of the accumulator by \code{CMPXCHG}, which has
no branchless form because a 32-bit write also clears the upper
half of the register.

As we saw above, the branchless code often looks quite different
from the original, and it is hard to verify the correctness of the refactoring
by manual review.
We therefore took a formal approach here. Sail's SMT backend
can check whether two pieces of Sail code return the same result
for every input. For each rewrite, we used it to prove that the
code before and after the rewrite agree. We proved 509 such
equivalences, one or more for each of the 138 rewrites, counting
the width and lane variants of a rewrite as one, almost all when
the rewrite was made and a few added afterwards. The properties state the
helper expressions before and after the change rather than
importing the instruction handlers, so they must be kept in step
with the code.

To measure the effect of the refactoring, we ran the 6,192
straight-line instruction sequences of Section~\ref{sec:applications} through
both the current model and the model before the branch removal.
As the baseline, we applied Isla's linearization to every function
of the old model that the pass accepts, and we gave each execution
60 seconds. Table~\ref{tab:branch-removal} shows the outcomes. The
current model finished 5,707 sequences within the limit (92\%), the
old model 4,726 (76\%), and the old model fails in ways the
refactored one does not. It runs out of time more often. A bit
range at a data dependent position means that a bit vector is
indexed by a variable, which Isla cannot handle. A vector index
out of bounds arises because Isla's linearization evaluates both
sides of a branch: it evaluates, for example, \code{GPR[idx - 4]}
on the side the original code would not take, which causes an error
when \code{idx} is 0. And even when it finishes within the limit,
it often produces several execution paths instead of one, so an
equivalence query built on Isla's output has to cover all of them.
The remaining sequences fail on both models for the same
reasons, mostly floating-point arithmetic that the model leaves to
external functions. Section~\ref{sec:applications} checks the
sequences that the current model finishes on a single path. A finished execution
takes 0.12 seconds at the median on the old model and 0.28 on the
current one: the refactoring costs a little on the easy sequences
and makes the hard ones tractable.

\begin{table}[t]
\caption{Symbolic execution of the 6,192 straight-line instruction
sequences of Section~\ref{sec:applications} on the model before
and after the branch removal, with a limit of 60 seconds per
sequence.}
\label{tab:branch-removal}
\centering
\begin{tabular}{@{}lrr@{}}
\toprule
Outcome & Before & After \\
\midrule
Finished on a single path & 2,224 & 5,442 \\
Finished on several paths & 2,502 & 265 \\
Exceeded the time limit & 511 & 154 \\
Bit range at a data dependent position & 427 & 0 \\
Vector index out of bounds & 111 & 0 \\
\bottomrule
\end{tabular}
\end{table}

While working on this refactoring, we found a bug in Isla's expression
simplifier that produced a false counterexample for two equivalent
instruction sequences. We submitted a fix
upstream.\anon[]{\footnote{\url{https://github.com/rems-project/isla/pull/106}}}

\section{Verifying Assembly Optimizations}
\label{sec:applications}

As an application of \proj, we asked an LLM to optimize the
handwritten assembly of seven widely used programs and checked
every proposal with the model. These projects use assembly for
performance, so we expected the code to be already well
optimized. Finding better sequences and automatically proving them
equivalent to the originals would demonstrate a practical use of
\proj, and every proposal that the check refutes shows why the
check is needed.

\subsection{Blocks from Production Code}
\label{sec:blocks}

We took all the handwritten x86-64 assembly of the seven programs
in Table~\ref{tab:corpus}, assembled as their build systems
assemble it, and extracted from the object code every
straight-line block. A block is a maximal run of 4 to 60
instructions with no control transfer and no branch target. We
excluded instructions that our equivalence check cannot handle,
such as stack operations and floating-point arithmetic.
A block may read and
write memory through up to three base registers, each with an
index register and displacements that fit a 256-byte window, and
may load the constants that the code keeps in its read-only data,
such as rounding terms, shuffle masks, and multipliers. Blocks that
differ only in register names are kept once. This gives 9,313
blocks, 6,192 of them distinct, from 377,716 lines of assembly.

\begin{table}[t]
\caption{The assembly corpus: files of handwritten assembly, their
lines, and the distinct straight-line blocks extracted from them.}
\label{tab:corpus}
\centering
\begin{tabular}{@{}lrrr@{}}
\toprule
Program & Files & Lines & Blocks \\
\midrule
FFmpeg (libav*/x86) & 153 & 67,932 & 1,861 \\
libjpeg-turbo (simd/x86\_64) & 27 & 9,494 & 90 \\
dav1d (src/x86) & 46 & 160,750 & 2,025 \\
Linux (crypto and CRC routines) & 52 & 29,965 & 556 \\
OpenSSL (perlasm output) & 25 & 77,578 & 382 \\
glibc (string, wcsmbs) & 100 & 25,723 & 997 \\
GMP (mpn/x86\_64) & 46 & 6,274 & 281 \\
\midrule
Total & 449 & 377,716 & 6,192 \\
\bottomrule
\end{tabular}
\end{table}

For each block we computed the registers and flags live at its
end by a backward analysis over the function's control-flow graph,
conservative in the direction that matters: an instruction the
analysis does not classify is assumed to read every register it
names, a call to read everything, and only a write of a whole
register kills it. The live set is what a rewrite must preserve;
every other register may be clobbered.

We also had to be careful not to require a newer processor than
the original code targets. Replacing SSE2 code with AVX2
instructions, for example, is not considered an optimization,
because the routine is likely to exist
for machines that have only SSE2. The projects make the target
visible in their file names in most cases, such as \code{\_sse2.s}
or \code{\_avx2.s}, or in their function names, and we restricted
each rewrite to the instruction set that the original block was
written for.

\subsection{Proposals and Checks}
\label{sec:checking}

We gave each block to the LLM with its ISA level, the registers
used in addresses, the constants, and the live registers and
flags, and asked for a sequence with fewer instructions that leaves
the same values in the live registers, performs the same stores,
and loads only from the same addresses.
The LLM reasoned without running code.
We assembled each proposal with a directive that makes the
assembler reject anything newer than the block's ISA level.

To check a proposal, both the original and the proposal
run symbolically from the same
initial state: every register symbolic, except that base registers
point into the memory window, index registers hold fixed small
values, and the constants are placed in the window; the window's
other bytes are symbolic and shared by both sequences. Isla
produces one path per sequence with the branch-free model, and we
ask an SMT solver for an input on which the live registers, the
live flags, or the window differ after the two sequences.
\code{unsat} proves that the rewrite produces the same results as
the original; \code{sat}
gives a counterexample. We gave each symbolic execution and each
solver call 60 seconds; a check that exceeds either limit counts
as unresolved.

\subsection{Results}
\label{sec:rewrites}

Table~\ref{tab:rewrite-results} gives the outcome per program. Of
the 6,192 blocks, the model executes 5,446 symbolically within the
limit. We asked GPT-6 Astra to propose a shorter sequence for each
of them. It proposed one for 789 blocks and declared the rest
already tight. The checks proved 714 of the proposals equivalent
to their originals, refuted 41
with a counterexample, and left 29 unresolved within the limits;
5 did not assemble. The
proved rewrites remove 1,436 of the 12,186 instructions in their
blocks.

\begin{table*}[t]
\caption{Rewrites of straight-line blocks proposed by the LLM and
checked with \proj. ``Checkable'' blocks are those the model
executes symbolically on a single path within the limit;
``unresolved'' includes the 5 proposals that did not assemble.}
\label{tab:rewrite-results}
\centering
\begin{tabular}{@{}lrrrrrr@{}}
\toprule
Program & Blocks & Checkable & Proposed & Proved & Refuted & Unresolved \\
\midrule
FFmpeg & 1,861 & 1,507 & 289 & 266 & 12 & 11 \\
libjpeg-turbo & 90 & 84 & 28 & 28 & 0 & 0 \\
dav1d & 2,025 & 1,909 & 197 & 173 & 16 & 8 \\
Linux & 556 & 378 & 58 & 45 & 5 & 8 \\
OpenSSL & 382 & 323 & 50 & 37 & 6 & 7 \\
glibc & 997 & 989 & 145 & 143 & 2 & 0 \\
GMP & 281 & 256 & 22 & 22 & 0 & 0 \\
\midrule
Total & 6,192 & 5,446 & 789 & 714 & 41 & 34 \\
\bottomrule
\end{tabular}
\end{table*}

The proved rewrites fall into three kinds. In 63, the LLM only
deleted instructions that the block did not need, such as a copy
whose destination the rest of the block never reads or a redundant
zeroing. In 168, it kept the block's set of instructions but used
fewer of them, most often by folding a load into the memory
operand of the instruction that consumed it. The remaining 483
introduce new instructions: \code{LEA} takes over an add-and-shift
chain, or a \code{PSHUFD} replaces a copy and an unpack.
Of the 41 refuted proposals, 39 deleted an instruction that
changes the stack pointer, such as an \code{ADD RSP, 32} after a
call or a \code{MOV RSP, RBP} that restores it. The block itself
never reads the stack pointer again, so the LLM took it for dead,
but the code after the block depends on it, and the check always
compares it.

To illustrate how an LLM rewrites assembly code and how \proj
checks the results, we present three examples: a proved rewrite,
a refuted proposal, and a case where the solver timed out. These
proposals came from a separate experiment but were generated in
a similar way.

\subsection{A Proved Rewrite}
\label{sec:upsample}

One rewrite we proved correct comes from libjpeg-turbo's chroma
upsampler (\code{jdsample-sse2.asm}), which doubles the horizontal
resolution of a row by interpolation. For each sample $s$ with
neighbors $l$ and $r$, it outputs
\[
  \lfloor (3s + l + 1)/4 \rfloor
  \quad\text{and}\quad
  \lfloor (3s + r + 2)/4 \rfloor.
\]
The original code widens each 8-bit sample to 16 bits, multiplies
by three, adds the neighbor and the constant 1 or 2, shifts, and
packs the results back into bytes, 27 instructions in all. Every
step is done twice, once for each half of the sixteen samples.

The LLM's rewrite stays in bytes and uses \code{PAVGB}, which
computes $\lfloor (a+b+1)/2 \rfloor$, an average that rounds up.
Averaging $s$ with a neighbor and then with $s$ again weights $s$
three to one, which is the interpolation the upsampler wants.
The trick is that $a + b - \lfloor (a+b+1)/2 \rfloor$ is
$\lfloor (a+b)/2 \rfloor$, an average that rounds down, and that
this still holds when the byte addition and subtraction wrap at 256.
Rounding up in the inner average
and down in the outer one gives the constant 1, and the reverse
gives 2:
\begin{align*}
  \bigl\lfloor \bigl(s + \lfloor (s+l+1)/2 \rfloor\bigr)/2 \bigr\rfloor
    &= \lfloor (3s+l+1)/4 \rfloor, \\
  \bigl\lfloor \bigl(s + \lfloor (s+r)/2 \rfloor + 1\bigr)/2 \bigr\rfloor
    &= \lfloor (3s+r+2)/4 \rfloor.
\end{align*}
The rewritten sequence, 13 instructions, needs no widening and no
constants, and works on all sixteen samples at once.

Convincing oneself that this transformation is correct takes some
thought. The equivalence check with \proj settled the question
automatically. Its \code{unsat} result establishes that the
rewrite computes the same result as the original for all
possible inputs.

\subsection{A Counterexample}
\label{sec:counterexamples}

One routine in FFmpeg's blend filter computes
$\lfloor a \times b / 255 \rfloor$ where $a$ and $b$ are 8-bit
values. The original code computes
\[
  f(x)=\bigl(x+1+((x+1)\mathbin{\gg}8)\bigr)
        \mathbin{\gg}8,
\]
where $x = a \times b$ is a 16-bit value.
The LLM proposed
\[
  g(x)=\bigl((x+128)\times257\bigr)\mathbin{\gg}16
\]
as a rewrite. This is a common idiom for division by 255,
so it looks plausible.
The equivalence check using \proj nevertheless rejected it.
The difference is that the original rounds down and the
replacement rounds to nearest.
For example, the inputs $a=128$ and $b=1$ give $x=128$, for which
the original returns 0 and the replacement returns 1.

\subsection{An Unresolved Rewrite}

One of the unresolved cases comes from the AES-GCM code in Linux's crypto
subsystem. It involves multiplication in GF($2^{128}$) using
\code{PCLMULQDQ}, which multiplies polynomials over GF(2).
Both operands are symbolic, yielding a large AND-XOR circuit.
Equivalence checking of such arithmetic circuits is known to be
very hard for SAT solvers~\cite{lv2013galois}.

\section{Architectural Findings from Validation}
\label{sec:manual}

Most discrepancies exposed by validation were caused by errors in our translation.
In several cases, however, validation revealed limitations or differences in the architectural documentation itself, including incorrect pseudocode, differences between vendor specifications, and instructions or encodings omitted from the Instruction Set Reference volume of the Intel SDM.

\subsection{A Zero-Count Shift Still Writes the Destination}

Consider \code{SHL EAX, CL} in 64-bit mode. The
instruction shifts EAX to the left by the count in CL and
writes the result back to EAX. In x86-64, a write to a
32-bit register clears the upper 32 bits of the 64-bit register it
aliases, so after this instruction the upper half of RAX must be zero.

Our first Sail model, however, did not clear the upper half when
CL was zero. For an initial RAX of
0x0123456789ABCDEF and CL=0, the results were:
\par\noindent\begin{minipage}{\linewidth}
\begin{lstlisting}
  hardware: 0x0000000089ABCDEF
  model:    0x0123456789ABCDEF
\end{lstlisting}
\end{minipage}\par
The cause of this difference is the SDM's pseudocode for \code{SHL}, which writes no
register when the count is zero. A faithful translation therefore
performs no write to EAX, so the upper 32 bits of RAX are not
cleared either. We changed the model
to match the hardware and reported it to Intel as a documentation issue.
Intel corrected the pseudocode in the June 2026 edition of the SDM~\cite{intelsdm092}.

\subsection{Differences Between the Intel and AMD Manuals}
\label{sec:bsf}

\code{BSF} (Bit Scan Forward) and \code{BSR} (Bit Scan Reverse) find the positions of the least and most significant set bits in their source operands, respectively.
When the source operand is zero, Intel's manual states that the destination is undefined.
Our initial model therefore represented the destination as undefined, and the generated emulator chose zero.
However, this behavior prevented Linux from booting.

The Linux kernel contains inline assembly that depends on the destination being preserved when the source operand is zero~\cite{linuxbitops}.
AMD's manual specifies this behavior~\cite{amdapm3}, and both our tested Intel and AMD processors exhibit it.
We therefore model destination preservation, while noting that this choice is stronger than the guarantee stated in Intel's manual.
This case shows that a practical x86-64 specification must reconcile vendor documentation, observed hardware behavior, and software dependencies rather than treating a single manual as authoritative.

\subsection{Obsolete x87 Instructions and Encodings}
\label{sec:x87-aliases}

The Instruction Set Reference volume of the Intel SDM is intended to list every instruction, but it omits several obsolete x87 instructions and encodings.
They appear only in the System Programming Guide under the heading ``Obsolete Instructions and Undefined Opcodes.''
The omissions include three obsolete x87 instructions that both tested processors execute as no-ops and 72 additional obsolete encodings for x87 operations.

Hardware differential testing exposed their absence from the model: the emulator raised an undefined-opcode exception, whereas both processors accepted them.
We added the missing instructions and encodings to the model.
Because these forms do not appear in the Instruction Set Reference, they are not included in the 4,218 instruction forms used for the coverage calculation.

\section{Related Work}

\paragraph{Formalizing x86}
The K semantics of Dasgupta et al.~\cite{dasgupta2019complete}
were validated with hardware execution, compiler tests, and
comparisons with other semantics. That work also found errors in
Intel's manual. The x86isa model lives inside the ACL2 theorem
prover, which makes it both executable and a basis for
interactive proofs about programs, including system programs
that manipulate page tables~\cite{goel2017engineering,goel2017books}.
Such proofs can cover loops and unbounded inputs, which our
bounded SMT checks cannot. Its current source includes support
for running a modified Linux kernel~\cite{x86isa2026source}, and
x86isa has been translated into Sail~\cite{taylor2026sailx86}.
Table~\ref{tab:scope} compares their architectural scope with \proj.

\paragraph{Building models from documentation}
Blanqui et al.~\cite{blanqui2011designing} generate an ARMv6
simulator and a Coq specification from the reference manual using
custom extraction tools and manually supplied constraints.
Booting Linux exposed both a pseudocode error and an overlooked
exception rule. Reid~\cite{reid2016trustworthy} derives executable
Armv8-A and Armv8-M specifications from the same sources used to
produce Arm's architecture manuals and validates them against
test suites, simulators, and processor designs. Our work uses
LLM-assisted translation to construct an x86-64 Sail model from
Intel's manual.

\paragraph{Learning behavior from processors}
Strata~\cite{heule2016stratified} takes a list of instructions and
an assembler for them as given, and automatically
discovers what the instructions do. It starts with a few dozen
instructions whose semantics were written by hand. For every
other instruction, it runs the instruction on the processor with
many random inputs, searches for a short sequence of instructions
with known semantics that produces the same outputs, and adopts
that sequence as the instruction's semantics. Each instruction
learned this way can then be used to learn others.

libLISA~\cite{craaijo2024liblisa} starts with almost no prior
knowledge of the architecture. It discovers instruction encodings
by executing byte sequences on the processor, then synthesizes
their semantics from the observed inputs and outputs.
These methods capture how particular processors behave, including
choices that the manual leaves undefined. Execution alone cannot
distinguish architectural guarantees from implementation choices.
We start from the documentation and use hardware execution to
check its translation.

\paragraph{Validating models and implementations}
Earlier work differentially tested CPU emulators and virtual CPUs against physical processors or reference simulators.
EmuFuzzer~\cite{10.1145/1572272.1572303} compares user-mode CPU emulators with physical processors, KEmuFuzzer~\cite{10.1145/1831708.1831730} extends such testing to privileged execution, and Amit et al.~\cite{10.1145/2815400.2815420} adapt Intel's CPU validation infrastructure to virtual CPUs, comparing their behavior against Intel's architectural simulator, which served as the reference in the absence of a complete formal model of the CPU.
Reid~\cite{reid2017guards} looks for bugs in Arm's formal
architecture specification itself, validating the Arm v8-M ASL
specification against independently stated architectural properties
and finding twelve defects, including two security relevant bugs.
InSPECtor~\cite{chesser2026inspector} checks processor specifications
against real hardware by generating tests from SLEIGH, a language
for describing instruction decoding and semantics.
EXAMINER~\cite{jiang2022examiner} uses symbolic execution of Arm's
ASL specification to generate instruction streams and compares
their behavior on CPU emulators and real processors.
DiffSpec~\cite{rao2024diffspec} uses LLMs to generate small test
programs from specifications and source code, runs each test on
multiple eBPF or WebAssembly implementations, and reports
behavioral discrepancies as potential bugs.
We validate \proj through instruction-level comparisons with real
processors, existing test suites, and system-level execution
including booting operating systems.

\paragraph{Tools and applications}
Sail and Isla provide the specification language and symbolic
execution infrastructure used in our
work~\cite{armstrong2019isa,armstrong2021isla}.
Islaris~\cite{sammler2022islaris} builds on them to verify functional
properties of machine-code programs against Sail ISA semantics
using program logic. Our use of Isla is narrower: we use it to
check equivalence between short instruction sequences.
Pydrofoil~\cite{bolz2025pydrofoil} accelerates Sail simulators through
specialized optimizations and a tracing JIT. Our compiler instead
lowers Sail models to LLVM for ahead-of-time compilation. We compare
it with the official Sail compiler, but not with Pydrofoil.

Separating candidate generation from correctness checking is
established practice in superoptimization~\cite{massalin1987,schkufza2013stochastic}.
LLM-Vectorizer~\cite{taneja2025vectorizer} uses LLMs to generate
vectorized code and Alive2 to check its correctness at the LLVM IR
level. We apply the same separation to handwritten assembly: an
LLM proposes rewrites, and Isla uses \proj to check whether the
original and rewritten machine-code sequences agree on selected
outputs under the stated assumptions. Our experiments demonstrate
an application of the generated ISA model rather than a new
superoptimization algorithm. Register availability, ISA
requirements, and the surrounding code must still be checked
before applying a rewrite.

\section{Discussion and Limitations}

Our experience suggests that no single source is sufficient to
validate a large ISA specification. Manuals describe architectural
intent but can contain errors or ambiguities; hardware provides an
independent oracle but may expose implementation choices rather
than architectural guarantees; and system execution exercises
interactions that isolated instruction tests often miss. We
therefore relied on these sources together.

An executable ISA specification also serves purposes that place
different demands on its representation. Code that closely follows
a manual is easier to review against the documentation, while code
intended for symbolic execution may need a substantially different
structure. Our experience suggests that these goals need not be
satisfied by the same representation: a relatively direct
specification can first be validated and then transformed into an
analysis-oriented form, with equivalence checking used to preserve
its semantics.

\proj nevertheless has important limitations. Floating-point
arithmetic lies outside the Sail semantics, multiprocessor
execution requires a separate memory model that we have not
provided, and several architectural extensions listed in
\S\ref{sec:scope} remain unsupported. The roughly four-week
development time should be interpreted as a case study rather than
a controlled measurement of LLM productivity. Although the
construction and validation process is not inherently specific to
x86-64, we have not evaluated it on another architecture.

\section{Conclusion}

We presented \proj, a comprehensive formal specification of
x86-64. With substantial LLM assistance, we constructed a Sail
model that covers both the instruction set and major system
mechanisms. We validated it using manual review, hardware
differential testing, and system execution, booting unmodified
Linux, FreeBSD, FreeDOS, Windows 3.1, and Windows 95 and running a guest through Linux's
KVM driver. This process also exposed a bug in the Intel SDM's
pseudocode and a case where Linux relies on behavior that the
manual leaves undefined.

We further transformed the specification into a form suitable for
symbolic execution using equivalence-checked refactoring. With the
resulting model, we checked the LLM's proposals for shorter
versions of 5,446 blocks of handwritten assembly from seven
programs, proving 714 rewrites and rejecting 41 with concrete
counterexamples. Together, these
results show that a comprehensive executable ISA specification can
support both full-system execution and formal reasoning about
production machine code.

\bibliographystyle{ACM-Reference-Format}
\bibliography{references}

\end{document}